\documentclass[namedreferences]{solarphysics}

\usepackage[hyperref,optionalrh]{spr-sola-addons} 
\usepackage{graphicx}        
\usepackage{color}           

\ifx \urlurl    \undefined \def \urlurl#1{\href{http://#1}{\textsf{#1}}}\fi
\ifx \doiurl    \undefined \def \doiurl#1{\href{http://dx.doi.org/#1}{\textsf{DOI}}}\fi
\ifx \adsurl    \undefined \def \adsurl#1{\href{http://adsabs.harvard.edu/abs/#1}{\textsf{ADS}}}\fi
\ifx \arxivurl  \undefined \def \arxivurl#1{\href{http://arxiv.org/abs/#1}{\textsf{arXiv}}}\fi

\chardef\us=`\_

\begin{document}

\begin{article}
\begin{opening}

\title{A Statistical Study of Distant Consequences of Large Solar Energetic Events}

\author{Carolus J.\ \surname{Schrijver}$^1$ \sep Paul A.\ \surname{Higgins}$^{1,2}$}

\runningtitle{Distant Consequences of Large Solar Energetic Events}
\runningauthor{C.J.\ Schrijver, P.A.\ Higgins}

\institute{$^1$ Lockheed Martin Advanced Technology Center,
               3251 Hanover St., Palo Alto, USA, 
                     email: \url{schrijver@lmsal.com}, \\
               $^2$ School of Physics, Trinity College Dublin, College Green, Dublin, Ireland}


\begin{abstract}
  Large solar flares and eruptions may influence remote regions
  through perturbations in the outer-atmospheric magnetic field,
  leading to causally related events outside of the primary or
  triggering eruptions that are referred to as ``sympathetic
  events''. We quantify the occurrence of sympathetic events using the
  full-disk observations by the {\it Atmospheric Imaging Assembly} onboard
  the {\it Solar Dynamics Observatory} associated with all flares of
  GOES
  class M5 or larger from 01 May 2010 through 31 December 2014. Using a
  superposed-epoch analysis, we find an increase in the rate of
  flares, filament eruptions, and substantial sprays and surges more
  than 20$^\circ$ away from the primary flares within the first four\,hours
  at a significance of 1.8 standard deviations. We also find that the
  rate of distant events drops by two standard deviations, or a factor
  of 1.2, when comparing intervals between 4\,hours and 24\,hours before and
  after the start times of the primary large flares. We discuss the
  evidence for the concluding hypothesis that the gradual evolution
  leading to the large flare and the impulsive release of the energy
  in that flare both contribute to the destabilization of magnetic
  configurations in distant active regions and quiet-Sun areas. These
  effects appear to leave distant regions, in an ensemble sense, in a
  more stable state, so that fewer energetic events happen for at
  least a day following large energetic events.
\end{abstract}

\end{opening}

\section{Introduction}
\label{sec:introduction}
Over the past decades, full-disk chromospheric and coronal imagers
have revealed a great deal of activity in the solar outer atmosphere,
with much attention going to the energetic phenomena of flares and
eruptions. Evidence is accumulating rapidly that coronal perturbations
have a long reach across the Sun as instrument cadence, sensitivity,
and wavelength coverage increase, in particular in recent years with
the deployment of the {\it Atmospheric Imaging Assembly}
\citep[AIA:][]{aiainstrument} onboard the {\it Solar Dynamics Observatory}
\citep[SDO:][]{2012SoPh..275....3P}. This includes the perturbations and
deformations of the coronal magnetic field associated with the initial
phases of coronal mass ejections (CMEs), involving various wave and
wave-like processes ({\it e.g.} \citealp{2012ApJ...750..134D,
  2012ApJ...753...52L, 2012SoPh..281..187P, 2013ApJ...776...58N,
  2014SoPh..289.3233L}, and references therein) and the slower
deformation of high magnetic field either by distant flux emergence or
by nearby eruptions ({\it e.g.} \citealp{schrijver+title2010,
  2013ApJ...773...93S}, and references therein).  With the advances
in observational capabilities, together with increasing computational
tools, answers to questions about couplings between energetic solar
events may be coming within reach ({\it e.g.} \citealp{2011ApJ...739L..63T,
  2012SoPh..280..389J, 2013ApJ...764...87L}), but it remains a
challenge to observers to establish the frequency at which significant
long-range interactions occur.

Among the interactions considered, one commonly referred to as
``sympathy'' has received particular attention, because it relates
both to the understanding of what is involved in the destabilization
of magnetic configurations and to the development of heliospheric
events such as CMEs and solar energetic particle (SEP) events (see
\citealp{2012ApJ...750...45H}, for an extensive discussion on a sample
set of interacting events).  ``Sympathy'' ---~defined here as the
coupling of events in distinct solar regions through the modification
of the atmospheric magnetic field~--- can occur either by the gradual
evolution of the surface magnetic field (through emergence,
displacement, or cancellation) or by relatively sudden changes that
are associated with explosive and/or eruptive events. The interactions
induced by the gradual changes associated with flux emergence have
been studied, for example by \citeauthor{2015arXiv150406633F}
(\citeyear{2015arXiv150406633F}, and references therein). They
conclude that ``newly emerging regions produce a significant increase
in the occurrence rate of X- and M-class flares in pre-existing
regions'' other than close to, or at, the location of new flux
emergence (where they note the effect is significant at about the
level of two standard deviations in the event counts, assuming Poisson
statistics). In this case, the stored (free) energy of the large flare
is released from an active region as its field configuration is
altered by the gradual evolution of one or more regions external to
the destabilized field, either by induction alone or by the eventual
reconnection-enabled field reconfiguration. Such couplings are, of
course, not limited to active regions, but may -- as noted by
\cite{2015arXiv150406633F} -- also affect quiet-Sun configurations
including large filaments and the CMEs that their destabilization may
induce.

On shorter time scales, eruption-related disturbances in the coronal
field may be involved in sympathy. These disturbances may be transient
as in the case of waves, or irreversible as may happen during large
coronal mass ejections, or a mixture of both. One example, studied
both on the Sun and in virtual settings, is the coupling between
adjacent quiet-Sun filaments, where the eruption of one plausibly
leads to the eruption of another one ({\it e.g.}
\citealp{schrijver+title2010,2011ApJ...739L..63T}).

The full-Sun, high-cadence view of the solar corona offered by SDO/AIA
has shown many examples in which couplings between events could be
inferred to exist, either through waves or to lasting field
perturbations.  A sample of such events was described in detail by
\cite{2013ApJ...773...93S}. However, although the couplings in many cases
appear plausible, the statistical evidence about the frequency and
impact of such couplings needed work. \cite{2015arXiv150406633F} now
provide such information on couplings between flux emergence in one
location and flaring in another. In this study, we look at couplings
between large impulsive events (associated with flares of magnitude M5
or larger) in one location and any substantial energetic event
elsewhere on the visible hemisphere on the Sun.

\begin{figure}
\centering
\includegraphics[width=\textwidth]{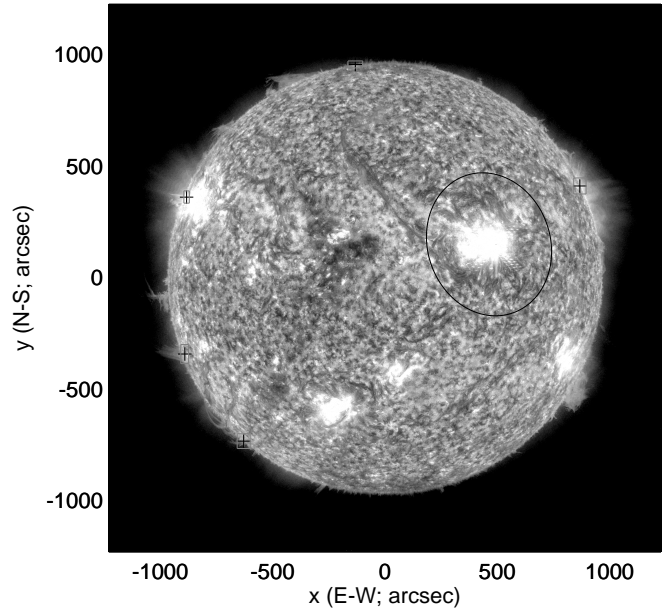}
\caption{Image taken by SDO/AIA in the 304\,\AA\ passband on 07
  September 22:35\,UT, exposing during an X1.8 flare
  (SOL2011-09-07T22:32:32L223C076). A distance from the flare's center
  of 20 heliocentric degrees is shown by the black circle projected
  onto the Sun as viewed from SDO's perspective. Five other
  events were flagged within 24\,hours of this flare, shown by pluses
  and their bounding boxes in grey, in this case all near or at the
  limb.}
\label{fig:addedfigure}
\end{figure}
\section{Observations and Results}
\label{sec:observations}
We start from the list of flares of GOES class M5 or larger that occurred
within 70$^\circ$ from disk center in the period of
post-commissioning SDO observations (starting on 01 May 2010)
through 31 December 2014. For
each flare, we inspected SDO/AIA 304\,\AA\ image sequences for flares,
filament eruptions, and large surges/sprays that occurred anywhere on
the disk and over the limb for periods spanning at least 24\,hours before
and after the start times of each selected large flare. For each such identified event, we
recorded the central position, bounding box, estimated start time, and
whether the event was a flare and/or eruption, and whether the event
occurred in active- or quiet-Sun settings. We also inspected
image sequences of the {\it Large Angle and Spectrometric COronagraph}
(\citealp[LASCO:][]{brueckner+etal1995}) on the {\it SOlar and Heliospheric Observatory} (SOHO)
image sequences to assess
whether the large flares appeared to be associated with a CME.

\nocite{2013ApJ...776...58N}
\begin{table}
  \caption{Selected flares and association with a CME
    (in SOHO/LASCO data) differentiated the number of events flagged in
    304\,\AA\ data over 20$^\circ$ from the flare site, respectively,
    within 24\,hours of flare start times. Column 3: plane of the sky velocities
    [$v_{\rm C}$] of
    CME. Column 4: velocities [$V_{\rm L}$] of large-scale
    propagating fronts (LCPF; from Nitta {\it et al.}, 2013), with values for events outside of the
    interval in the original study shown between brackets as determined
    later with the same procedure, or dots if no LCPF is observed. If
    a distant AIA event 
    occurs within four hours following the flare, an asterisk is added to
    the last column.}\label{tab:data}
\begin{tabular}{crrrr@{ CME }l}
\hline
SOL & Class & $v_{\rm [}$(km\,s$^{-1}$] & $v_{\rm L}$ [km\,s$^{-1}$]     & \multicolumn{2}{c}{Notes} \\
\hline
\multicolumn{6}{c}{ 18 flares with no distant events within 24\,hours:}\\
SOL2011-07-30T02:03:44L001C090 & M9.3 &  - & 383 & no & \\
SOL2011-08-03T13:16:48L333C074 & M6.0 & 610  & 604 & & \\
SOL2011-08-04T03:40:48L331C071 & M9.3 & 1315  & 910 & & \\
SOL2011-08-09T07:48:16L296C073 & X6.9 &  1610 & 743 & & \\
SOL2012-03-05T02:30:24L302C073 & X1.1 &  594 & 915 & & \\
SOL2012-03-07T00:02:08L329C090 & X5.4 &  1825 & 828 & & \\
SOL2012-03-07T01:05:04L317C068 & X1.3 &  `` & 789 & &(?) \\
SOL2012-03-10T17:14:40L280C090 & M8.4 &  491 & 522 & & \\
SOL2012-07-04T09:46:40L209C110 & M5.3 &  - & - & no & \\
SOL2012-10-23T03:13:04L167C090 & X1.8 &  - & - & no & \\
SOL2012-11-13T01:58:24L252C090 & M6.0 &  611 & - & & \\
SOL2013-10-25T07:53:36L001C090 & X1.7 &  587 & [973] & & \\
SOL2013-10-25T14:51:44L358C090 & X2.1 &  1081 & [1314] & & \\
SOL2013-11-05T22:06:56L162C102 & X3.3 &  562 & \ldots & & \\
SOL2014-01-07T10:08:00L093C103 & M7.2 &  451 & \ldots & & \\
SOL2014-01-30T15:47:12L100C103 & M6.6 &  780 & \ldots & & \\
SOL2014-03-29T17:36:00L145C079 & X1.0 &  528 & [1561] & & \\
SOL2014-04-02T13:17:52L010C076 & M6.5 & 1471 & [703] & & \\
\hline
\multicolumn{6}{c}{ 13 flares with one distant event within 24\,hours:}\\
SOL2011-09-06T22:12:16L227C076 & X2.1 &  575 & 1246 & & *\\
SOL2011-09-08T15:32:16L226C076 & M6.7 &  214 & 649 & no(?) & \\
SOL2011-09-24T09:21:04L278C078 & X1.9 &  1936 & 1129 & & \\
SOL2011-09-24T12:33:04L337C090 & M7.1 &  1915 & 640 & & *\\
SOL2012-05-10T04:10:40L179C077 & M5.7 &  - & - & no & *\\
SOL2012-07-05T11:38:40L108C112 & M6.1 &  - & - & no & \\
SOL2012-07-06T23:01:20L156C090 & X1.1 &  1828 & - & & \\
SOL2012-07-12T15:36:32L082C105 & X1.4 &  843 & 542 & & *\\
SOL2013-10-24T00:21:20L010C100 & M9.3 &  - & [594] & no & *\\
SOL2013-10-28T01:40:16L031C086 & X1.0 &  695 & [1403] & & \\
SOL2013-11-08T04:20:16L163C104 & X1.1 &  497 & \ldots & & *\\
SOL2014-01-07T18:03:44L100C090 & X1.2 &  1830 & [1036] & & *\\
SOL2014-02-04T03:56:48L105C104 & M5.2 &  - & \ldots & no & *\\
\hline
\multicolumn{6}{c}{ 16 flares with two or more distant events within 24\,hours:}\\
SOL2011-09-06T01:34:56L227C076 & M5.3 &  782 & 1041 & & *\\
SOL2011-09-07T22:32:32L223C076 & X1.8 &  792 & 1307 & & *\\
SOL2011-09-24T20:28:48L332C090 & M5.8 &  - & - & no & *\\
SOL2011-09-25T04:30:56L281C079 & M7.4 &  788 & 740 & & *\\
SOL2011-11-03T20:16:00L101C068 & X1.9 &  - & - & no & *\\
SOL2012-01-23T03:37:36L208C062 & M8.7 &  2175 & 837 & &\ \ \\
SOL2012-03-09T03:21:36L301C090 & M6.3 &  950 & 689 & & *\\
SOL2012-03-13T17:12:32L241C090 & M7.9 &  1884 & 1022 & & *\\
SOL2012-07-02T10:43:12L209C107 & M5.6 & ? & 1234 & & *\\
SOL2012-07-28T20:43:44L173C115 & M6.1 &  420 & [546]  & & *\\
SOL2013-04-11T06:56:00L073C081 & M6.5 &  861 & [719] & & *\\
SOL2013-05-15T01:25:20L296C078 & X1.2 &  1366 & \ldots & &  *\\
SOL2013-11-01T19:46:08L262C101 & M6.3 &  268 & \ldots & & *\\
SOL2013-11-10T05:07:12L164C104 & X1.1 & 682 & \ldots & & *\\
SOL2013-12-31T21:45:36L226C106 & M6.4 &  271 & \ldots & & *\\
SOL2014-01-01T18:40:00L227C104 & M9.9 &  236 & \ldots & & \\
\hline
\end{tabular}
\end{table}

The selected reference events are listed in Table~\ref{tab:data},
identified using the IAU SOL target naming convention \citep{2010SoPh..263....1.}. In
order to quantify the frequency of long-range couplings, we identified
other flares and eruptions well away from the flaring active
region. We set a threshold of 20$^\circ$ for the minimum distance
between the flare site and the central position of any other
event (see Figure~\ref{fig:addedfigure} for an example of this). 
The results are sorted chronologically within each of three
groups in Table~\ref{tab:data}. The first set contains the 18 flares
with no other events identified elsewhere on the Sun at least
$20^\circ$ from the flare site as seen by SDO/AIA within 24\,hours on
either side of the flare. The second set, with 13 flares, contains
those flares with a single other event meeting the above criteria. The
third set of 16 flares lists those with two or more events flagged.

\begin{figure}
\centering
\includegraphics[width=\textwidth]{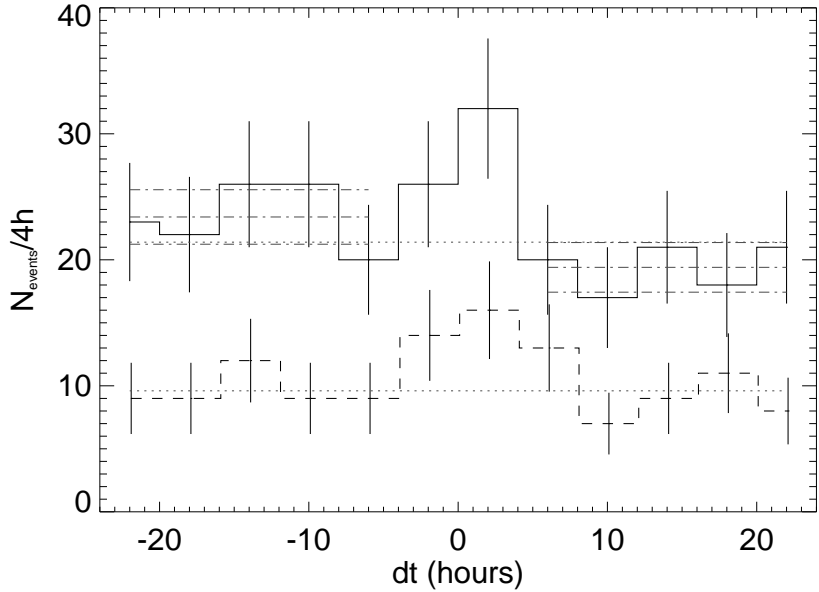}
\caption{Superposed-epoch histogram of the number of solar events (per
  four-hour interval) around the time of flares of GOES class M5 or larger
  in the period from 01 May 2010 through 31 December 2014 that occurred
  within 70 heliocentric degrees from disk center. The solid
  histograms shows the number distribution for all events (flares and
  eruptions) on the visible hemisphere or beyond the limb with central
  coordinates more than 20 heliocentric degrees away from the flare
  location. The dashed histogram shows the same but for all events
  originating in quiet-Sun regions. The uncertainty bars (slightly
  offset for the quiet-Sun events to avoid overlap) assume Poisson
  statistics. The dotted lines show the average event rates for the
  two samples over the 32\,hour intervals excluding times within four hours of
  the flare peak times. The grey dash--dotted lines show the averages
  (and the one $\sigma$ ranges) from four hours to 24\,hours before and after the
  flare start times, respectively.}
\label{fig:results}
\end{figure}
Figure~\ref{fig:results} is a superposed epoch rendering of the
frequency of observed events as function of time difference.  It
summarizes the distribution of the number of flagged events that
occurred more than 20$^\circ$ away from major flares. The diagram
shows event counts in four-hour intervals within 24\,hours of the reference
times that are defined as the start times the GOES X-ray flares. The
distributions suggest that there is an increased number of events in
the four-hour interval 
immediately following large flares at about the 1.8\ $\sigma$ level; for
the set of events occurring in quiet-Sun regions, a comparable
difference is seen at about 1.6\,$\sigma$ (assuming Poisson
statistics). For the full set of events, the mean event rate following
flares appears to be somewhat lower well after the flares ($4-24$\,hours)
than in the corresponding interval before; the difference of about 1.7
standard deviations equals a ratio of $1.2 \pm 0.2$.

\begin{figure}
\centering
\includegraphics[width=0.5\textwidth]{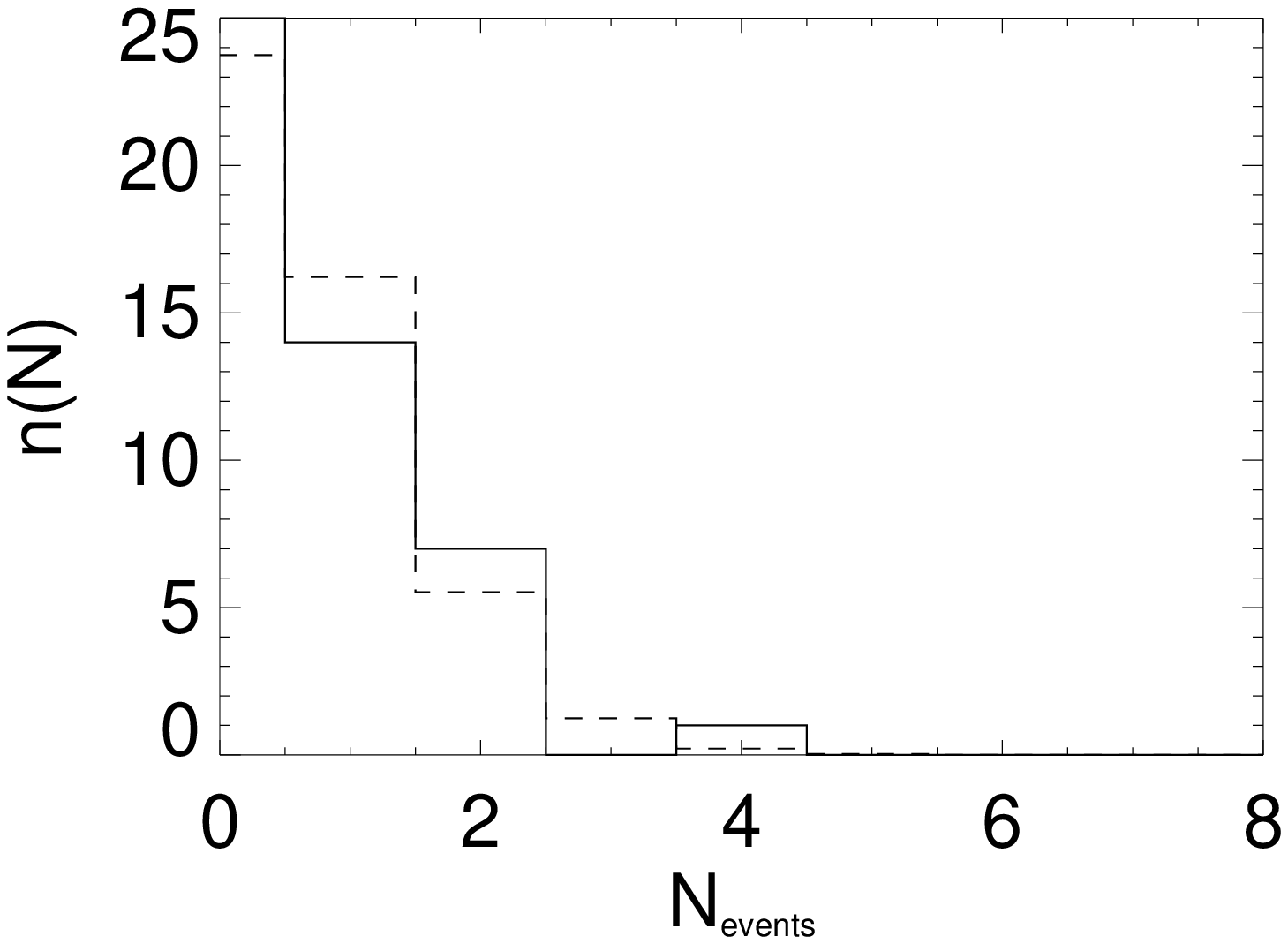}\includegraphics[width=0.5\textwidth]{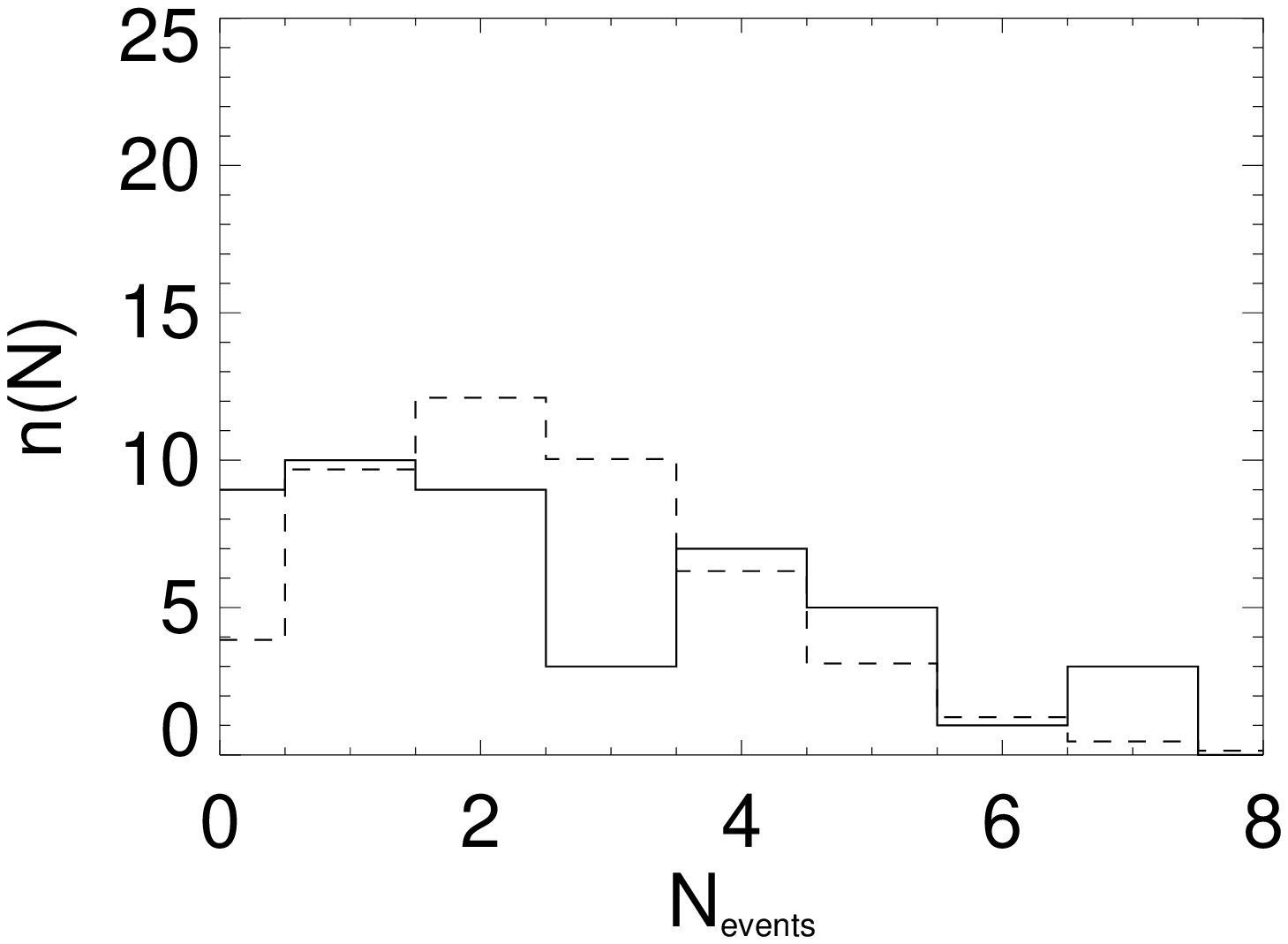}
\caption{{(left)} Histogram of the number of events observed in
  the SDO/AIA 304\,\AA\ image sequences within four hours following, and
  more than 20$^\circ$ from, each selected M5-class flare or larger
  (solid). For comparison, a Poisson distribution with the same
  average number of events (0.7) is shown (dashed). {(right)}
  Same, but for events between 24 and 4\,hours before the flares, and a
  Poisson distribution with average of 1.5 events.}
\label{fig:histograms}
\end{figure}
Figure~\ref{fig:histograms} shows the distribution of the number of
events within four hours following each large flare (solid) compared to a Poisson
distribution of the same mean number of events (dashed). The second panel shows
the same for an interval between 16\,hours and 4\,hours prior to the flare start time. Both cases
are roughly compatible with the assumption of Poisson statistics.
We may conclude that the majority of events present random waiting
times, but we learn little from that in the context of this study.

The data summarized in Table~\ref{tab:data} show that the fraction of
M5 flares or larger with CMEs is 0.8, 0.6, and 0.8 for flares with zero,
one, or more than one distant events within 24\,hours of the flare peak time,
{\it i.e.} there is no significant distinction in CME fraction as a function
of the number of distant events.  Average flare classes (using an averaging based on the
values of the flare peak brightness as used in flare class, and specifying
the standard deviation in the mean) are X (1.7$\pm 0.4$), X (1.1$\pm
0.1$), and M (9$\pm 1$). Again, there is no significant difference in flare class
given the number of distant events, particularly
given the range of flare classes in each group.

The average ``linear speeds'' (from
\urlurl{cdaw.gsfc.nasa.gov/CME\_list/}) and the associated
standard deviations in the mean for CMEs are: $890 \pm 130$\,km\,s$^{-1}$, $1150 \pm
250$\,km\,s$^{-1}$, and $880 \pm 170$\,km\,s$^{-1}$, respectively. There is no obvious
trend in these averages as function of the number of distant events
within four hours from the flare start times, but ``linear speed'' is a
plane-of-the-sky quantity and thus only in part informative about the
actual propagation speeds of CMEs.
 
More informative about what happens in the lower corona may be the
occurrence and properties of large-scale propagating fronts
(LCPFs) that are associated with many eruptive
events. Table~\ref{tab:data} lists the velocities measured by
\cite{2013ApJ...776...58N} for events between April 2010 and January
2013, and some later events measured by N.\,Nitta (private
communication, 2015) with the same algorithm: for zero, one, and two or more
distant events within a four-hour time window, these velocities and the
standard deviations in the mean are $854\pm 98$\,km\,s$^{-1}$, $905\pm
127$\,km\,s$^{-1}$, and $903\pm 92$\,km\,s$^{-1}$, respectively. We thus find no
significant trend in the number of ``sympathetic events'' with
velocity of the LCPF.

\section{Discussion}
\label{sec:discussion}
The review of full-disk SDO/AIA 304\,\AA\ observations of 48\,hour
intervals centered on flares of class M5 or larger reveals that other
flares, filament eruptions, and large sprays/surges that occur more
than 20$^\circ$ away from the reference flare site have a marginally
significant increase in their rate during the first four hours following the
large reference flares. This interval is followed by a period of at
least 20\,hours during which the event rates elsewhere on the disk are
likely to be somewhat reduced relative to the rate $4-24$\,hours before
the large reference flares.

Owing to the relatively weak solar cycle, the number of large flares was
low, contributing to the fact that these results are statistically
significant only at about the 1.7 to 1.8 standard deviation level. In view of
the other evidence in support of ``sympathy'' between
solar energetic events as discussed in Section~\ref{sec:introduction},
we note that the results from this study are consistent with the
hypothesis that major flares may cause the destabilization of distant
magnetic configurations. We consider it likely that the large flares
and associated coronal perturbations do this by causing 
destabilization to occur earlier than would occur in the absence of such large
flaring, {\it i.e.} by affecting primarily the timing of these events, and
possibly also their occurrence in the first place.

The data not only show a moderately significant increase in
distant--event rates within the first four hours following major
flares, but also suggest a drop by about two standard deviations, or
by a factor of about $1.2 \pm 0.2$, in distant--event rates when
comparing intervals between four and 24\,hours before and after the
flares. We propose that this drop is associated in large part with the
effects of flux emergence within the flaring region: Large flares are
often associated with signatures of flux emergence within a
pre-existing active region ({\it e.g.} \citealp{schrijver2007,
  welsch+li2008, schrijver2009b, 2009ApJ...705..821W}), and flux
emergence enhances the probability of flaring in distant regions
(\citealp{2015arXiv150406633F}). The decrease by a factor of about 1.2
in event rates between four and 24\,hours before and after M5 flares
or larger is roughly comparable to the relative change in flare
frequencies (also M5 or larger) between three-day intervals before and
after flux emergence of a factor of $1.2 - 1.7$ found by
\cite{2015arXiv150406633F}, being larger for samples restricted to
larger emerging flux regions.

In conclusion, we propose a scenario with two components: the gradual
evolution leading to large flares and the impulsive release of stored
energy in the flare process jointly contribute to the destabilization
of magnetic configurations, both in distant active regions and in
quiet-Sun areas. The large impulsive events can lead to
destabilization of relatively distant regions, leaving these in a
lower energy or more stable state so that fewer energetic events
happen for at least a day following the large primary events. 
Part of the drop in the rate
of distant events before and after major flaring may also reflect that
major flares occur towards the end of the emergence of a flux rope
within active regions, but that aspect remains to be tested.

\acknowledgements{This work was supported by NASA 's SDO/AIA contract
  (NNG04EA00C) to LMSAL. AIA is an instrument onboard the {\it Solar
  Dynamics Observatory}, a mission for NASA's Living With a Star
  program.
Data are provided courtesy of NASA/SDO and the AIA science team.
We thank George Lee for his help in
  reviewing and annotating the AIA observations, and Nariaki Nitta for
  applying his perturbation-tracking algorithm to some of the more
  recent large-scale propagating fronts.}

\section*{Disclosure of Potential Conflicts of Interest}
The authors declare that they have no conflicts of interest.


\begin{thebibliography}{20}
\ifx\bisbn     \undefined \def\bisbn  #1{ISBN #1}\fi
\ifx\binits    \undefined \def\binits#1{#1}\fi
\ifx\bauthor   \undefined \def\bauthor#1{#1}\fi
\ifx\batitle   \undefined \def\batitle#1{#1}\fi
\ifx\bjtitle   \undefined \def\bjtitle#1{\textit{#1}}\fi
\ifx\bvolume   \undefined \def\bvolume#1{\textbf{#1}}\fi
\ifx\byear     \undefined \def\byear#1{#1}\fi
\ifx\bissue    \undefined \def\bissue#1{#1}\fi
\ifx\bfpage    \undefined \def\bfpage#1{#1}\fi
\ifx\blpage    \undefined \def\blpage #1{#1}\fi
\ifx\burl      \undefined \def\burl#1{\textsf{#1}}\fi
\ifx\href      \undefined \def\href#1#2{\textsf{#2}}\fi
\ifx\betal     \undefined \def\betal{\textit{et al.}}\fi
\ifx\bctitle   \undefined \def\bctitle#1{#1}\fi
\ifx\beditor   \undefined \def\beditor#1{#1}\fi
\ifx\bbtitle   \undefined \def\bbtitle#1{\textit{#1}}\fi
\ifx\bedition  \undefined \def\bedition#1{#1}\fi
\ifx\bseriesno \undefined \def\bseriesno#1{\textbf{#1}}\fi
\ifx\blocation \undefined \def\blocation#1{#1}\fi
\ifx\bsertitle \undefined \def\bsertitle#1{\textit{#1}}\fi
\ifx\bsnm      \undefined \def\bsnm#1{#1}\fi
\ifx\bsuffix   \undefined \def\bsuffix#1{#1}\fi
\ifx\bparticle \undefined \def\bparticle#1{#1}\fi
\ifx\barticle  \undefined \def\barticle#1{}\fi
\ifx\binstitute  \undefined \def\binstitute#1{#1}\fi
\ifx\bpublisher  \undefined \def\bpublisher#1{#1}\fi
\ifx\doiurl    \undefined
  \def\doiurl#1{\href{http://dx.doi.org/#1}{\textsf{DOI}}}\fi
\ifx\arxivurl  \undefined
  \def\arxivurl#1{\href{http://arxiv.org/abs/#1}{\textsf{arXiv}}}\fi
\ifx\adsurl    \undefined
  \def\adsurl#1{\href{http://adsabs.harvard.edu/abs/#1}{\textsf{ADS}}}\fi
\ifx\botherref \undefined \def\botherref#1{}\fi
\ifx\url       \undefined \def\url#1{\textsf{#1}}\fi
\ifx\bchapter  \undefined \def\bchapter#1{}\fi
\ifx\bbook     \undefined \def\bbook#1{}\fi
\ifx\bcomment  \undefined \def\bcomment#1{#1}\fi
\ifx\oauthor   \undefined \def\oauthor#1{#1}\fi
\ifx\citeauthoryear \undefined\def \citeauthoryear#1{#1}\fi
\ifx\endbibitem\undefined \def\endbibitem{}\fi
\ifx\bconflocation  \undefined \def\bconflocation#1{#1} \fi

\bibitem[\protect\citeauthoryear{{Brueckner}
  \textit{et~al.}}{1995}]{brueckner+etal1995}
\begin{barticle}
\bauthor{\bsnm{{Brueckner}}, \binits{G.E.}},
\bauthor{\bsnm{{Howard}}, \binits{R.A.}},
\bauthor{\bsnm{{Koomen}}, \binits{M.J.}},
\bauthor{\bsnm{{Korendyke}}, \binits{C.M.}},
\bauthor{\bsnm{{Michels}}, \binits{D.J.}},
\bauthor{\bsnm{{Moses}}, \binits{J.D.}},
\bauthor{\bsnm{{Socker}}, \binits{D.G.}},
\bauthor{\bsnm{{Dere}}, \binits{K.P.}},
\bauthor{\bsnm{{Lamy}}, \binits{P.L.}},
\bauthor{\bsnm{{Llebaria}}, \binits{A.}},
\bauthor{\bsnm{{Bout}}, \binits{M.V.}},
\bauthor{\bsnm{{Schwenn}}, \binits{R.}},
\bauthor{\bsnm{{Simnett}}, \binits{G.M.}},
\bauthor{\bsnm{{Bedford}}, \binits{D.K.}},
\bauthor{\bsnm{{Eyles}}, \binits{C.J.}}:
\byear{1995},
\batitle{{The Large Angle Spectroscopic Coronagraph (LASCO)}}.
\bjtitle{Solar Phys.}
\bvolume{162},
\bfpage{357}.
\doiurl{10.1007/BF00733434}.
\adsurl{1995SoPh..162..357B},
\end{barticle}
\endbibitem

\bibitem[\protect\citeauthoryear{{Downs}
  \textit{et~al.}}{2012}]{2012ApJ...750..134D}
\begin{barticle}
\bauthor{\bsnm{{Downs}}, \binits{C.}},
\bauthor{\bsnm{{Roussev}}, \binits{I.I.}},
\bauthor{\bsnm{{van der Holst}}, \binits{B.}},
\bauthor{\bsnm{{Lugaz}}, \binits{N.}},
\bauthor{\bsnm{{Sokolov}}, \binits{I.V.}}:
\byear{2012},
\batitle{{Understanding SDO/AIA Observations of the 2010 June 13 EUV Wave
  Event: Direct Insight from a Global Thermodynamic MHD Simulation}}.
\bjtitle{Astrophys. J.}
\bvolume{750},
\bfpage{134}.
\end{barticle}
\endbibitem

\bibitem[\protect\citeauthoryear{{Fu} and {Welsch}}{2015}]{2015arXiv150406633F}
\begin{botherref}
\oauthor{\bsnm{{Fu}}, \binits{Y.}},
\oauthor{\bsnm{{Welsch}}, \binits{B.T.}}:
2015,
{Triggering of Remote Flares by Magnetic Flux Emergence}.
\textit{ArXiv e-prints 1504.06633}.
\end{botherref}
\endbibitem

\bibitem[\protect\citeauthoryear{{Harrison}
  \textit{et~al.}}{2012}]{2012ApJ...750...45H}
\begin{barticle}
\bauthor{\bsnm{{Harrison}}, \binits{R.A.}},
\bauthor{\bsnm{{Davies}}, \binits{J.A.}},
\bauthor{\bsnm{{M{\"o}stl}}, \binits{C.}},
\bauthor{\bsnm{{Liu}}, \binits{Y.}},
\bauthor{\bsnm{{Temmer}}, \binits{M.}},
\bauthor{\bsnm{{Bisi}}, \binits{M.M.}},
\bauthor{\bsnm{{Eastwood}}, \binits{J.P.}},
\bauthor{\bsnm{{de Koning}}, \binits{C.A.}},
\bauthor{\bsnm{{Nitta}}, \binits{N.}},
\bauthor{\bsnm{{Rollett}}, \binits{T.}},
\bauthor{\bsnm{{Farrugia}}, \binits{C.J.}},
\bauthor{\bsnm{{Forsyth}}, \binits{R.J.}},
\bauthor{\bsnm{{Jackson}}, \binits{B.V.}},
\bauthor{\bsnm{{Jensen}}, \binits{E.A.}},
\bauthor{\bsnm{{Kilpua}}, \binits{E.K.J.}},
\bauthor{\bsnm{{Odstrcil}}, \binits{D.}},
\bauthor{\bsnm{{Webb}}, \binits{D.F.}}:
\byear{2012},
\batitle{{An Analysis of the Origin and Propagation of the Multiple Coronal
  Mass Ejections of 2010 August 1}}.
\bjtitle{Astrophys. J.}
\bvolume{750},
\bfpage{45}.
\end{barticle}
\endbibitem

\bibitem[\protect\citeauthoryear{{Jacobs} and
  {Poedts}}{2012}]{2012SoPh..280..389J}
\begin{barticle}
\bauthor{\bsnm{{Jacobs}}, \binits{C.}},
\bauthor{\bsnm{{Poedts}}, \binits{S.}}:
\byear{2012},
\batitle{{A Numerical Study of the Response of the Coronal Magnetic Field to
  Flux Emergence}}.
\bjtitle{Solar Phys.}
\bvolume{280},
\bfpage{389}.
\doiurl{10.1007/s11207-012-9941-8}.
\adsurl{2012SoPh..280..389J},
\end{barticle}
\endbibitem

\bibitem[\protect\citeauthoryear{Leibacher
  \textit{et~al.}}{2010}]{2010SoPh..263....1.}
\begin{barticle}
\bauthor{\bsnm{Leibacher}, \binits{J.}},
\bauthor{\bsnm{Sakurai}, \binits{T.}},
\bauthor{\bsnm{Schrijver}, \binits{C.J.}},
\bauthor{\bsnm{{van Driel-Gesztelyi}}, \binits{L.}}:
\byear{2010},
\batitle{{Solar Observation Target Identification Convention for use in Solar
  Physics}}.
\bjtitle{Solar Phys.}
\bvolume{263},
\bfpage{1}.
\doiurl{10.1007/s11207-010-9553-0}.
\adsurl{2010SoPh..263....1},
\end{barticle}
\endbibitem

\bibitem[\protect\citeauthoryear{{Lemen} \textit{et~al.}}{2012}]{aiainstrument}
\begin{barticle}
\bauthor{\bsnm{{Lemen}}, \binits{J.R.}},
\bauthor{\bsnm{{Title}}, \binits{A.M.}},
\bauthor{\bsnm{{Akin}}, \binits{D.J.}},
\bauthor{\bsnm{{Boerner}}, \binits{P.F.}},
\bauthor{\bsnm{{Chou}}, \binits{C.}},
\bauthor{\bsnm{{Drake}}, \binits{J.F.}},
\bauthor{\bsnm{{Duncan}}, \binits{D.W.}},
\bauthor{\bsnm{{Edwards}}, \binits{C.G.}},
\bauthor{\bsnm{{Friedlaender}}, \binits{F.M.}},
\bauthor{\bsnm{{Heyman}}, \binits{G.F.}},
\bauthor{\bsnm{{Hurlburt}}, \binits{N.E.}},
\bauthor{\bsnm{{Katz}}, \binits{N.L.}},
\bauthor{\bsnm{{Kushner}}, \binits{G.D.}},
\bauthor{\bsnm{{Levay}}, \binits{M.}},
\bauthor{\bsnm{{Lindgren}}, \binits{R.W.}},
\bauthor{\bsnm{{Mathur}}, \binits{D.P.}},
\bauthor{\bsnm{{McFeaters}}, \binits{E.L.}},
\bauthor{\bsnm{{Mitchell}}, \binits{S.}},
\bauthor{\bsnm{{Rehse}}, \binits{R.A.}},
\bauthor{\bsnm{{Schrijver}}, \binits{C.J.}},
\bauthor{\bsnm{{Springer}}, \binits{L.A.}},
\bauthor{\bsnm{{Stern}}, \binits{R.A.}},
\bauthor{\bsnm{{Tarbell}}, \binits{T.D.}},
\bauthor{\bsnm{{Wuelser}}, \binits{J.-P.}},
\bauthor{\bsnm{{Wolfson}}, \binits{C.J.}},
\bauthor{\bsnm{{Yanari}}, \binits{C.}},
\bauthor{\bsnm{{Bookbinder}}, \binits{J.A.}},
\bauthor{\bsnm{{Cheimets}}, \binits{P.N.}},
\bauthor{\bsnm{{Caldwell}}, \binits{D.}},
\bauthor{\bsnm{{Deluca}}, \binits{E.E.}},
\bauthor{\bsnm{{Gates}}, \binits{R.}},
\bauthor{\bsnm{{Golub}}, \binits{L.}},
\bauthor{\bsnm{{Park}}, \binits{S.}},
\bauthor{\bsnm{{Podgorski}}, \binits{W.A.}},
\bauthor{\bsnm{{Bush}}, \binits{R.I.}},
\bauthor{\bsnm{{Scherrer}}, \binits{P.H.}},
\bauthor{\bsnm{{Gummin}}, \binits{M.A.}},
\bauthor{\bsnm{{Smith}}, \binits{P.}},
\bauthor{\bsnm{{Auker}}, \binits{G.}},
\bauthor{\bsnm{{Jerram}}, \binits{P.}},
\bauthor{\bsnm{{Pool}}, \binits{P.}},
\bauthor{\bsnm{{Soufli}}, \binits{R.}},
\bauthor{\bsnm{{Windt}}, \binits{D.L.}},
\bauthor{\bsnm{{Beardsley}}, \binits{S.}},
\bauthor{\bsnm{{Clapp}}, \binits{M.}},
\bauthor{\bsnm{{Lang}}, \binits{J.}},
\bauthor{\bsnm{{Waltham}}, \binits{N.}}:
\byear{2012},
\batitle{{The Atmospheric Imaging Assembly (AIA) on the Solar Dynamics
  Observatory (SDO)}}.
\bjtitle{Solar Phys.}
\bvolume{275},
\bfpage{17}.
\doiurl{10.1007/s11207-011-9776-8}.
\adsurl{2012SoPh..275...17L},
\end{barticle}
\endbibitem

\bibitem[\protect\citeauthoryear{{Liu} and {Ofman}}{2014}]{2014SoPh..289.3233L}
\begin{barticle}
\bauthor{\bsnm{{Liu}}, \binits{W.}},
\bauthor{\bsnm{{Ofman}}, \binits{L.}}:
\byear{2014},
\batitle{{Advances in Observing Various Coronal EUV Waves in the SDO Era and
  Their Seismological Applications (Invited Review)}}.
\bjtitle{Solar Phys.}
\bvolume{289},
\bfpage{3233}.
\doiurl{s11207-014-0528-4}.
\adsurl{2014SoPh..289.3233L},
\end{barticle}
\endbibitem

\bibitem[\protect\citeauthoryear{{Liu}
  \textit{et~al.}}{2012}]{2012ApJ...753...52L}
\begin{barticle}
\bauthor{\bsnm{{Liu}}, \binits{W.}},
\bauthor{\bsnm{{Ofman}}, \binits{L.}},
\bauthor{\bsnm{{Nitta}}, \binits{N.V.}},
\bauthor{\bsnm{{Aschwanden}}, \binits{M.J.}},
\bauthor{\bsnm{{Schrijver}}, \binits{C.J.}},
\bauthor{\bsnm{{Title}}, \binits{A.M.}},
\bauthor{\bsnm{{Tarbell}}, \binits{T.D.}}:
\byear{2012},
\batitle{{Quasi-periodic Fast-mode Wave Trains within a Global EUV Wave and
  Sequential Transverse Oscillations Detected by SDO/AIA}}.
\bjtitle{Astrophys. J.}
\bvolume{753},
\bfpage{52}.
\end{barticle}
\endbibitem

\bibitem[\protect\citeauthoryear{{Lynch} and
  {Edmondson}}{2013}]{2013ApJ...764...87L}
\begin{barticle}
\bauthor{\bsnm{{Lynch}}, \binits{B.J.}},
\bauthor{\bsnm{{Edmondson}}, \binits{J.K.}}:
\byear{2013},
\batitle{{Sympathetic Magnetic Breakout Coronal Mass Ejections from
  Pseudostreamers}}.
\bjtitle{Astrophys. J.}
\bvolume{764},
\bfpage{87}.
\end{barticle}
\endbibitem

\bibitem[\protect\citeauthoryear{{Nitta}
  \textit{et~al.}}{2013}]{2013ApJ...776...58N}
\begin{barticle}
\bauthor{\bsnm{{Nitta}}, \binits{N.V.}},
\bauthor{\bsnm{{Schrijver}}, \binits{C.J.}},
\bauthor{\bsnm{{Title}}, \binits{A.M.}},
\bauthor{\bsnm{{Liu}}, \binits{W.}}:
\byear{2013},
\batitle{{Large-scale Coronal Propagating Fronts in Solar Eruptions as Observed
  by the Atmospheric Imaging Assembly on Board the Solar Dynamics Observatory -
  an Ensemble Study}}.
\bjtitle{Astrophys. J.}
\bvolume{776},
\bfpage{58}.
\doiurl{10.1007/s11207-012-0098-2}.
\adsurl{2012SoPh..281...67P},
\end{barticle}
\endbibitem

\bibitem[\protect\citeauthoryear{{Patsourakos} and
  {Vourlidas}}{2012}]{2012SoPh..281..187P}
\begin{barticle}
\bauthor{\bsnm{{Patsourakos}}, \binits{S.}},
\bauthor{\bsnm{{Vourlidas}}, \binits{A.}}:
\byear{2012},
\batitle{{On the Nature and Genesis of EUV Waves: A Synthesis of Observations
  from SOHO, STEREO, SDO, and Hinode (Invited Review)}}.
\bjtitle{Solar Phys.}
\bvolume{281},
\bfpage{187}.
\end{barticle}
\endbibitem

\bibitem[\protect\citeauthoryear{{Pesnell}, {Thompson}, and
  {Chamberlin}}{2012}]{2012SoPh..275....3P}
\begin{barticle}
\bauthor{\bsnm{{Pesnell}}, \binits{W.D.}},
\bauthor{\bsnm{{Thompson}}, \binits{B.J.}},
\bauthor{\bsnm{{Chamberlin}}, \binits{P.C.}}:
\byear{2012},
\batitle{{The Solar Dynamics Observatory (SDO)}}.
\bjtitle{Solar Phys.}
\bvolume{275},
\bfpage{3}.
\doiurl{10.1007/s11207-011-9841-3}.
\adsurl{2012SoPh..275....3P},
\end{barticle}
\endbibitem

\bibitem[\protect\citeauthoryear{{Schrijver}}{2007}]{schrijver2007}
\begin{barticle}
\bauthor{\bsnm{{Schrijver}}, \binits{C.J.}}:
\byear{2007},
\batitle{{A Characteristic Magnetic Field Pattern Associated with All Major
  Solar Flares and Its Use in Flare Forecasting}}.
\bjtitle{Astrophys. J. Lett.}
\bvolume{655},
\bfpage{117}.
\end{barticle}
\endbibitem

\bibitem[\protect\citeauthoryear{{Schrijver}}{2009}]{schrijver2009b}
\begin{barticle}
\bauthor{\bsnm{{Schrijver}}, \binits{C.J.}}:
\byear{2009},
\batitle{{Driving major solar flares and eruptions: A review}}.
\bjtitle{Advances in Space Research}
\bvolume{43},
\bfpage{739}.
\end{barticle}
\endbibitem

\bibitem[\protect\citeauthoryear{{Schrijver} and
  {Title}}{2011}]{schrijver+title2010}
\begin{barticle}
\bauthor{\bsnm{{Schrijver}}, \binits{C.J.}},
\bauthor{\bsnm{{Title}}, \binits{A.M.}}:
\byear{2011},
\batitle{{Long-range magnetic couplings between solar flares and coronal mass
  ejections observed by SDO and STEREO}}.
\bjtitle{Journal of Geophysical Research (Space Physics)}
\bvolume{116}(\bissue{A15}),
\bfpage{4108}.
\end{barticle}
\endbibitem

\bibitem[\protect\citeauthoryear{{Schrijver}
  \textit{et~al.}}{2013}]{2013ApJ...773...93S}
\begin{barticle}
\bauthor{\bsnm{{Schrijver}}, \binits{C.J.}},
\bauthor{\bsnm{{Title}}, \binits{A.M.}},
\bauthor{\bsnm{{Yeates}}, \binits{A.R.}},
\bauthor{\bsnm{{DeRosa}}, \binits{M.L.}}:
\byear{2013},
\batitle{{Pathways of Large-scale Magnetic Couplings between Solar Coronal
  Events}}.
\bjtitle{Astrophys. J.}
\bvolume{773},
\bfpage{93}.
\end{barticle}
\endbibitem

\bibitem[\protect\citeauthoryear{{T{\"o}r{\"o}k}
  \textit{et~al.}}{2011}]{2011ApJ...739L..63T}
\begin{barticle}
\bauthor{\bsnm{{T{\"o}r{\"o}k}}, \binits{T.}},
\bauthor{\bsnm{{Panasenco}}, \binits{O.}},
\bauthor{\bsnm{{Titov}}, \binits{V.S.}},
\bauthor{\bsnm{{Miki{\'c}}}, \binits{Z.}},
\bauthor{\bsnm{{Reeves}}, \binits{K.K.}},
\bauthor{\bsnm{{Velli}}, \binits{M.}},
\bauthor{\bsnm{{Linker}}, \binits{J.A.}},
\bauthor{\bsnm{{De Toma}}, \binits{G.}}:
\byear{2011},
\batitle{{A Model for Magnetically Coupled Sympathetic Eruptions}}.
\bjtitle{Astrophys. J., Lett.}
\bvolume{739},
\bfpage{L63}.
\end{barticle}
\endbibitem

\bibitem[\protect\citeauthoryear{{Welsch} and {Li}}{2008}]{welsch+li2008}
\begin{bchapter}
\bauthor{\bsnm{{Welsch}}, \binits{B.T.}},
\bauthor{\bsnm{{Li}}, \binits{Y.}}:
\byear{2008},
\bctitle{{On the Origin of Strong-Field Polarity Inversion Lines}}.
In: \beditor{\bsnm{{Howe}}, \binits{R.}},
\beditor{\bsnm{{Komm}}, \binits{R.W.}},
\beditor{\bsnm{{Balasubramaniam}}, \binits{K.S.}},
\beditor{\bsnm{{Petrie}}, \binits{G.J.D.}} (eds.)
\bbtitle{Subsurface and Atmospheric Influences on Solar Activity},
\bseriesno{CS-383},
Astron. Soc. Pacific, San Francisco,
\bfpage{429}.
\end{bchapter}
\endbibitem

\bibitem[\protect\citeauthoryear{{Welsch}
  \textit{et~al.}}{2009}]{2009ApJ...705..821W}
\begin{barticle}
\bauthor{\bsnm{{Welsch}}, \binits{B.T.}},
\bauthor{\bsnm{{Li}}, \binits{Y.}},
\bauthor{\bsnm{{Schuck}}, \binits{P.W.}},
\bauthor{\bsnm{{Fisher}}, \binits{G.H.}}:
\byear{2009},
\batitle{{What is the Relationship Between Photospheric Flow Fields and Solar
  Flares?}}
\bjtitle{Astrophys. J.}
\bvolume{705},
\bfpage{821}.
\end{barticle}
\endbibitem

\end{thebibliography}

\end{article}
\end{document}